\documentclass[preprint,eqseqnum,12pt]{elsarticle}
\usepackage{epsfig}
\usepackage{amssymb}
\usepackage{amsmath}
\usepackage{lineno}
\biboptions{sort&compress}

\usepackage{bm}% bold math
\usepackage{amsmath}
\usepackage[utf8]{inputenc}

\numberwithin{equation}{section}
\journal{}

\begin{document}

\begin{frontmatter}




\title{The Mass Gap Approach to QCD. \\ II. The non-perturbative renormalization theory for the massive full gluon propagator}

%\title{The Mass Gap Approach to QCD. \\ II. No mass-shell for the massive full gluon propagator}


\author{V. Gogokhia}
\ead{gogohia.vahtang@wigner.hu}
\author{G.G. Barnaf\"oldi}
\ead{barnafoldi.gergely@wigner.hu}

\address{Wigner Research Centre for Physics,\\

29-33 Konkoly-Thege Mikl\'os Str., H- 121, Budapest, Hungary}

\begin{abstract}
The previously formulated the mass gap approach to QCD, based on a new insights into its ground state (vacuum),  
makes it possible to exactly define the dynamically generated gluon pole mass.
We developed a non-perturbative multiplicative renormalization program for the corresponding massive full gluon propagator.
Its asymptotic properties have been analysed, including the perturbation theory limit to the free gluon state, i.e., no any asymptotic freedom effect.
The peculiarities of the mass-shell structure of the massive full gluon propagator has been discussed.
The inconsistency of the canonical gauge in QCD is fixed. Our approach does not allow the  massive gluons to be the mass-shell objects. This prevents them to appear
in the physical spectrum (confinement of massive gluon states). The massive gluons may exist in the vacuum or inside hadrons only. 
Expression for the massive full gluon propagator in Euclidean metric for the lattice simulations is present. It is also shown that the massive solution has a correct limit 
to the massless full gluon propagator, when the gluon pole mass is to be formally put zero.
\end{abstract}

%for gluons to acquire mass dynamically. 



\begin{keyword}
theoretical particle physics  \sep Quantum Chromodynamics  \sep non-perturbative renormalization \sep gauge \sep gluon mass \sep no mass-shell.
\PACS
11.10.-z \sep 11.15.-q \sep 12.38.-t \sep 12.38.Aw  \sep 12.38.Lg
\MSC[2020] 30D30 \sep 81Q40 \sep 81T13 \sep 81T16
\end{keyword}
\end{frontmatter}
%%\linenumbers

\hspace{2mm}

\section{Introduction}

\hspace{2mm}

In our previous work~\cite{1} (which is the first paper of the three paper series announced there) we have addressed the general problem of a mass dynamical/spontaneous generation
 in the QCD ground state (vacuum). 
The new significant insights in to its true dynamical and gauge structures has been obtained on the basis of the derivation of the exact constraint on any solution to QCD. 
This has been achieved by the self-consistent use of the Slavnov-Tyalor (ST) identities and the  corresponding dynamical equation of motion in the vacuum for the gauge particle, the so-called 
Schwinger-Dyson (SD) equation for the full gluon propagator~\cite{1,2,3,4} . In this way it was formulated the mass gap approach to  QCD and its true vacuum~\cite{1}.
It is worth noting that it has been done without making any kind of the simplifications and using the tensor algebra derivation rigorous rules only. 

Let us begin with the showing up the new expression for the regularized full gluon propagator $D_{\mu\nu}(q) =D_{\mu\nu}(q; D)$ in the covariant gauge, namely  

\begin{equation}
D_{\mu\nu}(q) =  - i T_{\mu\nu}(q){ 1 \over q^2 + q^2 \Pi(q^2) - M^2}  - i L_{\mu\nu}(q) { \lambda^{-1} \over q^2 - \lambda^{-1} M^2},
\end{equation}
where we have introduced the useful notation for the free gluon gauge-fixing parameter $\xi_0 = \lambda^{-1}$~\cite{5}.  
Contrary to the initial work~\cite{1}, this expression is given in Minkowski metric $q^2=q^2_0 - \bf q^2$ which is more convenient to study the massive gluon fields,
and the whole gluon momentum range is $q^2 \in [0, \infty)$. As usual $T_{\mu\nu}(q)  = g_{\mu\nu} -  L_{\mu\nu}(q) = g_{\mu\nu} -  q_{\mu}q_{\nu} / q^2$.
The invariant function $\Pi(q^2) = \Pi^s(q^2; D)$ of~\cite{1} is a regular at zero and only logarithmic divergent in the perturbation theory (PT) $q^2 \rightarrow \infty$ limit: otherwise 
it remains an arbitrary function of its variable $q^2$ (this notation is introduced in order to simplify further derived expressions). So it can be subject of the PT renormalization 
program (not our problem here). The only quadratically ultra-violet (UV) divergent constant, but regularized from above and below, present explicitly in eq.~(1.1), is the so-called tadpole term 
$\Delta^2_t(D)$ which has been already replaced by $M^2$, i.e., we put $\Delta^2_t(D) = M^2(D) = M^2$. It has the dimension of a mass squared at any $D$.
In~\cite{1} it has been exactly proven that it cannot be removed from the QCD and its ground state by any means.  Its renormalized version has been conventionally called a mass gap,
being responsible for the first confinement phase transition in QCD. 

From the full gluon propagator (1.1) the ST identity becomes
\begin{equation}
q_{\mu}q_{\nu} D_{\mu\nu}(q) = - i \xi(q^2)  = - i { \lambda^{-1} q^2 \over (q^2 - \lambda^{-1} M^2)},
\end{equation}
where $\xi(q^2) = \xi(q^2; D)$ is its gauge-fixing parameter.
We call $\xi(q^2)$ as the generalized gauge since it depends on $M^2$, and when it is formally put zero, one recovers the gauge-fixing parameter 
for the free massless gluon propagator. 
To our best knowledge the generalized gauge (1.2) has been derived for the first time in a such new manner in~\cite{1}.
In the generalized gauge the gauge-fixing parameter $\lambda^{-1}$ can vary continuously from zero to infinity.
The functional dependence of the generalized gauge-fixing parameter $\xi(q^2)$ (1.2) is fixed up to an arbitrary gauge-fixing parameter $\xi_0 = \lambda^{-1}$. Unless we fix it, and thus $\xi(q^2)$ 
itself, we will call such situation as the general gauge dependence (GGD), see eqs.~(1.1) and (1.2).
Choosing $\xi_0 = \lambda^{-1}$ explicitly, we will call such situation as the explicit gauge dependence (EGD).
For example, $\xi_0 = \lambda^{-1}=0$ is called the unitary (Landau) gauge, $\xi_0 = \lambda^{-1}=1$ is called the t' Hooft--Feynman gauge, etc.~\cite{6,7,8,9}. 
The formal $\xi_0 = \lambda^{-1}= \infty$ limit is called as the canonical gauge in~\cite{8}, and its properties
will be discussed in detail below. This distinction is not a mere convention in QCD because of the presence of the mass scale parameter in its ground state.
The generalized gauge directly follows from the GGD/EGD formalism within the mass gap approach to QCD. It requires that there is no other functional expression 
for $\xi$, apart from given by the relation (1.2) at finite  $\xi_0 = \lambda^{-1}$, in the full gluon propagator (1.1) for the regularized massive gluon fields.

\hspace{1mm}

\subsection{The self-consistency condition for the gauge choice \\ in the mass gap approach to QCD}

\hspace{1mm}

For the readers convenient, let us formulate the self-consistency condition for the gauge choice in QCD as follows:

\begin{equation}
{{\lambda}^{-1} q^2 \over q^2 -  {\lambda}^{-1} M^2} = {a q^2 \over q^2 -  a M^2},
\end{equation}
i.e., the left-hand-side of this equation is present by the generalized gauge expression (1.2), while its right-hand-side presents the same expression
when the gauge is already chosen. In other words, we are checking whether the GGD formalism is compatible with its EGD counterpart and $\it vice \ versa$. So that its aim is
to derive a relation (not an identity) involving the gauge-fixing parameter. It has been proposed for the first time in~\cite{1} and  is present above.
If $a$ is any finite number, then from the self-consistency relation (1.3) it is easy to derive that $ {\lambda}^{-1} = a$, indeed.
At the same time, if $a = \infty$ is the canonical gauge, mentioned above, then the relation (1.3) becomes

\begin{equation}
{{\lambda}^{-1} q^2 \over q^2 -  {\lambda}^{-1} M^2} = - {q^2 \over M^2},
\end{equation}
which is only satisfied at $q^2=0$, i.e., there is no any condition for the gauge-fixing parameter. These simple arguments point out that
something is really wrong with the canonical gauge ${\lambda}^{-1} = \infty$ in QCD, but this is not the whole story yet.
Let us remind~\cite{1} that the self-consistency condition has been formulated not as a some kind of the parametrization, but follows from the exact solutions of the ST identities.
As underlined above, they are important for the renormalizability of the theory. That is why this condition has a deep mathematical and physical meaning. So that the inconsistency of the canonical gauge   
is a physically meaningful. In other words, if the condition (1.3) has a real solution then the theory is renormalizable, if not then the theory is not renormalizable (see below for the detailed discussion).

To formulate and perform the non-perturbative (NP) multiplicative (MP) renormalization program for the regularized full gluon propagator (1.1), obtained within the  
mass gap approach to QCD and its true vacuum,  in the case of massive gluons, will be the main purpose of this work.

\section{NP MP renormalization of the massive full gluon propagator}

Let us now perform the renormalization program for the full gluon propagator, presented in the relations (1.1) and (1.2).
If the denominator in eq.~(1.1) may have a zero at the some finite point $q^2=M_g^2$, then one obtains
\begin{equation}
M^2 =[ 1  +  \Pi(M^2_g)]M^2_g = {\tilde Z}_3^{-1} M^2_g,
\end{equation}
where
\begin{equation}
{\tilde Z}_3^{-1} \equiv {\tilde Z}_3^{-1}(M^2_g)= [ 1  +  \Pi(M^2_g)],
\end{equation}
and it can be treated as the mass gap NP MP renormalization constant, while $M^2$ being the 'bare' gluon mass, and $M^2_g$
being its renormalized counterpart, i.e., the mass gap itself, as pointed out above. It can be interpreted as a gluon pole mass, since
it is exactly defined by the relation (2.1). Due to the non-linear (NL) transcendental character of the denominator in the full gluon propagator (1.1), the arbitrary invariant function 
$\Pi(q^2)$ at the finite point $q^2=M_g^2$ may be indeed quadratically divergent, when the UV regulating parameter goes to infinity.

Let us further expand the invariant function $\Pi(q^2)$ around the pole position, namely
\begin{equation}
\Pi(q^2) = \Pi(M^2_g) + (q^2 - M^2_g) \Pi'(M^2_g) + (q^2 - M^2_g)^2 \Pi''(M^2_g) + ...
\end{equation}
This expansion is the corresponding Taylor series from the mathematical point of view, and thus is convergent, but from the physical point of view it is the so-called ''cluster'' expansion, since the dependence
of its coefficients on the coupling constant is not known~\cite{10}. The transcendental NL equation, which is the denominator in eq.~(1.1), has been equivalently replaced by 
the above-mentioned cluster expansion, see eq.~(2.4) below. It is much more convenient for the formulation of the NP renormalization program.

Substituting all these expressions back into the denominator and doing some algebra, one obtains
\begin{equation}
q^2 + q^2 \Pi(q^2) - M^2 = (q^2 - M^2_g) Z_3^{-1} [ 1 + Z_3 {\tilde \Pi}(q^2)],
\end{equation}
where
\begin{equation}
Z_3^{-1} \equiv Z_3^{-1}(M^2_g) = 1 + \Pi(M^2_g) + M^2_g \Pi'(M^2_g) = {\tilde Z}_3^{-1} + M^2_g \Pi'(M^2_g),
\end{equation}
and it can be treated as the additional NP MP renormalization constant for the transverse part of the full gluon propagator. 
The invariant function ${\tilde \Pi}(q^2)$ is defined as follows:
\begin{equation}
{\tilde \Pi}(q^2) = (q^2 - M^2_g) \left[ \Pi'(M^2_g) + q^2 \Pi''(M^2_g) + q^2(q^2 - M^2_g) \Pi'''(M^2_g) + ... \right].
\end{equation}
In the derivation of the relations (2.4)-(2.6) we have used the following obvious identity $q^2 = (M_g^2 + q^2 - M^2_g)$ in the term
with the first derivative $\Pi'(M^2_g)$. Let us point out that the renormalized invariant function, ${\tilde \Pi}(q^2)$ in eq.~(2.6) has the following
interesting properties, namely
\begin{equation}
{\tilde \Pi}(M^2_g)=0, \quad \quad {\tilde \Pi}(0) = - M^2_g \Pi'(M^2_g),
\end{equation}
so that it vanishes at the pole position $q^2=M^2_g$, and it is regular at zero.  

Substituting in its turn all these relations into 
eqs.~(1.1)-(1.2), one finally arrives at the following system for the
renormalized massive full gluon propagator $D^R_{\mu\nu}(q)$ within the GGD formalism, namely
\begin{equation}
D^R_{\mu\nu}(q) = { - i Z_3  {\tilde Z}^{-1}_3 \over (q^2 - M^2_g)[ 1 + Z_3 {\tilde \Pi}(q^2)]} T_{\mu\nu}(q)
- i L_{\mu\nu}(q)  {{\tilde\lambda}^{-1} \over (q^2 - {\tilde\lambda}^{-1} M_g^2)},
\end{equation}
where $D^R_{\mu\nu}(q)$ is defined as follows:
\begin{equation}
D^R_{\mu\nu}(q) = {\tilde Z}^{-1}_3 D_{\mu\nu}(q), \quad  \tilde\lambda = \lambda {\tilde Z}_3,
\quad Z_3 {\tilde Z}_3^{-1} = 1 - Z_3  M^2_g \Pi'(M^2_g).
\end{equation}
The renormalized ST identity becomes

\begin{equation}
q_{\mu}q_{\nu} D^R_{\mu\nu}(q)  = - i \xi^R(q^2) = - i {{\tilde\lambda}^{-1} q^2 \over q^2 -  {\tilde\lambda}^{-1} M_g^2}.
\end{equation}

From now on all the quadratically UV divergences disappear from the theory. Along with the arbitrary subtraction point they have been incorporated
into the NP MP renormalization
constants (2.2) and (2.5). The invariant function ${\tilde \Pi}(q^2)$ and the full gluon propagator itself are regular at zero. Thus, one can conclude
that the YM theory is NP renormalizable within our approach since all the quantities in the full gluon propagator (2.8), and hence in
the ST identity (2.10), are expressed in terms of the renormalized quantities only. The dependence on the NP MP renormalization constants
should disappear in QCD through the corresponding identity, which includes quark degrees of freedom. This is beyond the scope of the present investigation.

Neglecting the contribution from the regular part of the full gluon-self energy, i.e., putting formally
\begin{equation}
\Pi(q^2) = {\tilde \Pi}(q^2) = 0, \ Z_3 = {\tilde Z}_3 = 1, \ {\tilde\lambda} = \lambda, \ M^2_g = M^2,
\end{equation}
then from eq.~(2.8) for the massive free vector boson propagator in the generalized gauge, one obtains
\begin{equation}
D^0_{\mu\nu}(q; M^2_g) =  { - i \over (q^2 - M^2_g)} \left[ g_{\mu\nu} - (1 - {\tilde\lambda}^{-1})
{q_{\mu}q_{\nu} \over (q^2 - {\tilde\lambda}^{-1} M^2_g)} \right],
\end{equation}
which satisfies the same ST identity (2.10) as a function of ${\tilde\lambda}^{-1}$ and  $M^2_g$, as expected. This massive free vector particle propagator for the first time 
has been used in the investigations of the different models with spontaneously broken gauge theories in~\cite{6,7,8,9,10}. In~\cite{5} the last expression has been called as one 
being written down in the Stueckelberg gauge, so it naturally follows from the generalized gauge in eqs.~(2.8) and (2.10).

Within the EGD formalism let us show the free massive vector boson propagator (2.12) in some different finite gauges, such as
\begin{equation}
\textrm{unitary \ (Landau) \ gauge,} \ \  {\tilde\lambda}^{-1}=0: \quad  D^0_{\mu\nu}(q; M^2_g) = { - i T_{\mu\nu}(q) \over (q^2 - M^2_g)},
\end{equation}
\begin{equation}
\textrm{t'Hooft\,--\,Feynman \ gauge,} \ \ {\tilde\lambda}^{-1}=1:  \quad  D^0_{\mu\nu}(q; M^2_g) = { - i g_{\mu\nu} \over (q^2 - M^2_g)}.
\end{equation}

\section{Asymptotics of the massive full gluon propagator}

Let us now investigate the structure of the full gluon propagator with the finite gluon pole mass (2.8) in the $q^2 \rightarrow 0$ limit.
Then one arrives at
\begin{equation}
D^R_{\mu\nu}(q) = {  i {\tilde Z}^{-1}_3 \over M^2_g \left[ Z_3^{-1} + {\tilde \Pi}(0) \right]} T_{\mu\nu}(q) + i L_{\mu\nu} (q) { 1 \over M_g^2},
\quad q^2 \rightarrow 0
\end{equation}
Taking now into account relations (2.5) and (2.7), one finally obtains
\begin{equation}
iD^R_{\mu\nu}(0) = - { 1 \over M^2_g} T_{\mu\nu}(q) - L_{\mu\nu} (q) { 1 \over M_g^2} =  - { 1 \over M^2_g} g_{\mu\nu},
\end{equation}
and this result is exact, gauge-independent and regular at $q^2 = 0$. Evidently, this value coincides with the expression (2.12)
at any finite gauge in the $q^2 \rightarrow 0$ limit, as it needs be.

It is instructive to investigate the structure of the full gluon propagator with the gluon pole mass (2.8) in the PT $q^2 \rightarrow \infty$ limit,
but $M^2_g$ is finite. So that suppressing the term  $M^2_g/q^2$ in this limit in eq.~(2.8), one arrives at

\begin{equation}
D^R_{\mu\nu}(q) = { - i {\tilde Z}^{-1}_3 \over q^2 [ Z_3^{-1} + \tilde {\Pi}(q^2)]} T_{\mu\nu}(q)
-  {i \tilde {\lambda}^{-1} \over q^2} L_{\mu\nu}(q),
\end{equation}
where the expansion (2.6) now looks like

\begin{equation}
{\tilde \Pi}(q^2) = (q^2) \left[ \Pi'(M^2_g) + (q^2) \Pi''(M^2_g) + q^2(q^2) \Pi'''(M^2_g) + ... \right].
\end{equation}
%{\tilde \Pi}(q^2) = q^2 \Pi'(M^2_g) + q^4 \Pi''(M^2_g) + ˙q^6 \Pi'''(M^2_g) + ... .
Taking now in account expressions (2.2) and (2.5) as well as the definitions (2.9) and doing some algebra, one finally obtains

\begin{equation}
D_{\mu\nu}(q) = { - i  \over q^2 [ 1 + \Pi(q^2)]} T_{\mu\nu}(q) -  {i \lambda^{-1} \over q^2} L_{\mu\nu}(q),
\end{equation}
and the initial invariant function (2.3) now is given by the Taylor expansion as follows:

\begin{equation}
\Pi(q^2) = \sum_{n=0}^{\infty} (q^2)^n \Pi^{(n)}(M^2_g),
\end{equation}
in agreement with the expansion (2.3) in the PT limit at finite $M^2_g$, as expected, and where $\Pi^{(n)}(M^2_g)$ denotes its corresponding derivatives at the point $q^2=M^2_g$.
The expansion (3.6) is a formal one since it demonstrates an essential singularity
at large $q^2$. So that in this case the initial invariant function becomes some function having the essential singularity in the PT limit.
Then the corresponding effective charge looks like $d(q^2) = { 1 / 1 + \Pi(q^2)}, \quad q^2 \rightarrow \infty$.
The behaviour of any meromorphic function near its essential singularity is governed by the Picard theorem~\cite{11,12,13} (and see any textbook on the theory of functions of complex variable). 
It tells us that such kind of function in the close neighbourhood of its essential singularity can be replaced by
the constant $Z$, which value depends only on how precisely the essential singularity is achieving, i.e., $ \Pi(q^2) = Z, \quad q^2 \rightarrow \infty$.
The gluon momentum can go to infinity by the two different ways, namely as a loop variable or as the free particle momentum (i.e., not a loop variable). 
Putting $Z=0$, one obtains that the full gluon propagator (3.5) becomes the free gluon propagator in any case, namely

\begin{equation}
D^0_{\mu\nu}(q) = - i \left[ T_{\mu\nu}(q) +  \lambda^{-1} L_{\mu\nu}(q) \right] { 1 \over q^2},
\end{equation}
in complete agreement with eqs.~(2.11)-(2.12) in the PT $q^2 \rightarrow \infty$ limit.
Evidently, the massive full gluon propagator has no asymptotic freedom (AF)~\cite{14,15} structure in the PT limit, i.e., it does not satisfy the correct boundary condition~\cite{1} at high energies.

The important observation is that the dependence on the gluon pole mass $M^2_g$ vanishes in the massive full gluon propagator in the PT limit. 
It seems reasonable conclusion, since why indeed such defined mass
should survive this limit, i.e., even remaining finite it cannot determine the structure of the vacuum far away from the finite pole position.
In the third paper of our series we will show that the non-trivial PT dynamics 
in the YM vacuum is determined by the mass scale parameter, which distinguishes from $M^2_g$. Dynamically it is also generated by the tadpole term, but after performing 
the completely different renormalization program it becomes $\Lambda^2_{YM}$.

\section{Canonical gauge  ${\tilde\lambda}^{-1} = \infty$ }

Before going to investigate the most important issue of the massive full gluon propagator (2.8)'s structure on the mass-shell,
it is instructive to study the one particular gauge choice. It is in close connection with the gauge structure of the massive full gluon propagator (2.8)
on the mass-shell. Within the above-mentioned EGD formalism there exists one special gauge, the so-called canonical gauge, namely ${\tilde\lambda}^{-1} = \lambda^{-1} = \infty$. 
Then eq.~(2.8) becomes
\begin{equation}
D^R_{\mu\nu}(q) = { - i Z_3 {\tilde Z}^{-1}_3 \over (q^2 - M^2_g)[ 1 + Z_3 {\tilde \Pi}(q^2)]} T_{\mu\nu}(q)
+ i{q_{\mu}q_{\nu} \over q^2} {1 \over M_g^2}.
\end{equation}
Its ST identity now looks like
\begin{equation}
q_{\mu}q_{\nu} D^R_{\mu\nu}(q) = - i \xi^R(q^2) = i {q^2 \over M_g^2}.
\end{equation}

Neglecting now the contribution from the regular part of the full gluon self-energy in accordance with the relations (2.11), for the massive free 
counterpart in this gauge, one obtains
\begin{equation}
D^0_{\mu\nu}(q; M^2_g) = {- i \over (q^2 - M^2_g)} \left( g_{\mu\nu} - {q_{\mu}q_{\nu} \over M^2_g} \right) \sim \textrm{const.},  \quad q^2 \rightarrow \infty,
\end{equation}
which coincides with eq.~(2.12) in the $\lambda^{-1} = \infty$ limit, as expected, and the same ST identity, namely
\begin{equation}
q_{\mu}q_{\nu} D^0_{\mu\nu}(q; M^2_g)= -i \xi^R(q^2)  = i {q^2 \over M_g^2}.
\end{equation}

From the expressions (4.1)-(4.2) and clearly from eqs.~(4.3)-(4.4) it follows that the massive gluon propagator in this gauge has the
non-renormalizable behaviour in the PT $q^2 \rightarrow \infty$ limit, and the smooth $M^2_g=0$ limit does not exist as well.
In the quantum electroweak gauge theory such asymptotic, coming from the longitudinal component
of the propagators for massive vector particles $Z, W^+, W^-$ is not a problem. Due to the conserved currents in this theory, longitudinal components
of their propagators do not contribute to the $S$-matrix elements, describing this or that physical process/quantity.
In QCD such conserved currents do not exist, and moreover, the longitudinal components interact with each other~\cite{5}. The non-renormalizable behaviour
of the massive gluon propagator possesses a serious problem in this gauge. It has been called as the canonical one
in~\cite{8}, as mentioned above.

For further discussion, let us remind the self-consistency condition for the gauge choice in QCD. For the renormalized quantities it is given in the eqs.~(1.3) and (1.4) 
by the replacement there ${\lambda}^{-1}  \rightarrow {\tilde\lambda}^{-1}$ and $ M^2 \rightarrow  M_g^2$.
All this shows the general inconsistency of the canonical gauge ${\tilde\lambda}^{-1} = \infty$ in the mass gap approach to QCD. 
In other words, only finite gauges ${\tilde\lambda}^{-1}$ are compatible with it.
It formulates the interaction and gauge
of the massive full gluon propagator (2.8), associated with its transverse and longitudinal components, respectively, in such a way that the theory
remains renormalizable. The invariant function, associated with its
longitudinal component, is the known function of $q^2$ and  $M^2_g$. It explicitly demonstrates the equivalence between the PT $q^2 \rightarrow \infty$
and $M^2_g=0$ limits and $\it vice \ versa$, which is important for the renormalizability. The massive full gauge particle propagator (the corresponding Green's function) is at least quadratically convergent, i.e., behaves as $\sim (1/q^2)$ at $q^2 \rightarrow \infty$. In other words, our approach renders the YM massive theory renormalizable at infinity. Also, it provides the smooth transition $\xi \rightarrow \xi_0$ in the PT or $M^2_g=0$ limits.
However, the resulting (i.e., renormalized) massive gluon propagator in the canonical gauge ${\tilde\lambda}^{-1} = \infty$ makes the theory non-renormalizable, see  both eq.~(4.1) and eq.~(4.3).
Thus, the above-mentioned equivalence is broken down by this gauge. Therefore, the canonical gauge should
be abandoned, as not satisfying to the PT behaviour at infinity and having no smooth $M^2_g=0$ limit, both requested by the mass gap approach to QCD.

\section{No mass-shell for the massive full gluon propagator}

The one of the important issues of a mass dynamical generation in QCD is to be considered here. In order to put the discussion above on a firm mathematical ground by a few exact derivations,
let us begin with the structure of the renormalized massive full gluon propagator (2.8) on the mass-shell. In this section it will be temporally identified with the exactly defined gluon pole mass, namely
$q^2=M^2_g$ or simply mass-shell in what follows. Then one arrives at

\begin{equation}
D^R_{\mu\nu}(q) = { - i Z_3  {\tilde Z}^{-1}_3 \over (q^2 - M^2_g)} \left( g_{\mu\nu} - { q_{\mu}q_{\nu} \over M_g^2} \right)
- i \tilde{\epsilon} { q_{\mu}q_{\nu} \over M_g^4},
\end{equation}
because of the first of the relations (2.7) and where

\begin{equation}
\tilde{\epsilon} = { {\tilde\lambda}^{-1} \over (1 - {\tilde\lambda}^{-1})} =  { 1 \over ({\tilde\lambda} - 1)}.
\end{equation}
However, it is not difficult to show that eq.~(5.1) is equivalent to
\begin{equation}
D^R_{\mu\nu}(q) = { - i Z_3  {\tilde Z}^{-1}_3 \over (q^2 - M^2_g)} \left( g_{\mu\nu} - { q_{\mu}q_{\nu} \over M_g^2} \right)
\end{equation}
on the mass-shell at any $\tilde{\epsilon}$, and its free massive counterpart becomes

\begin{equation}
D^0_{\mu\nu}(q; M^2_g) = { - i \over (q^2 - M^2_g)} \left( g_{\mu\nu} - { q_{\mu}q_{\nu} \over M_g^2} \right),
\end{equation}
since in this case  $Z_3  {\tilde Z}^{-1}_3 = 1$ because of the relations (2.11). Evidently, eq.~(2.12) on the mass-shell coincides with the previous equation, as it needs be.
The corresponding ST identity looks like

\begin{equation}
q_{\mu}q_{\nu} D^0_{\mu\nu}(q; M^2_g)= -i \xi^R(q^2)  = i {q^2 \over M_g^2},
\end{equation}
which completely coincides with the ST identity (4.4) as need be.

Starting from the massive ful gluon propagator (2.8) in the arbitrary gauge, one concludes that its expression on the mass-shell coincides with the expressions written down in the canonical gauge  
(5.3)-(5.5), see section 4 as well. However, as we already know the canonical gauge ${\tilde\lambda}^{-1} = \infty$ is not self-consistent in the mass gap approach to QCD discussed in~\cite{1}
(see also the end of section 1). This means that the massive full gluon propagator (2.8) cannot be put on the mass-shell. In other words, the exactly defined gluon pole mass $q^2=M^2_g$ does not determine the massive full gluon propagator on the mass-shell, i.e., the mass-shell does not exist for the massive gluons though they may acquire mass dynamically, nevertheless. Therefore, there is no the dielectric 
phase transition for massive gluons, anyway~\cite{2} . So that they cannot appear as physical particles (confinement of the massive gluons within our approach). 

The regularization prescription $i0^+$ has been omitted in the massive gluon propagators. They are substantially modified due to the response of the true NP vacuum of QCD. This prescription is
designated for and can be applied to the theories with perturbative vacua~\cite{11,16}. Also, it is clear that we are working in the covariant gauges in order to avoid the peculiarities of the
non-covariant axial gauges~\cite{11} (and references therein).

\section{Conclusions}
%\label{sec:sum}

We have formulated and completed the NP MP renormalization program for the full gluon propagator (1.1) in the case of the massive gluons within the mass gap approach to QCD and its true ground state. 
On this way, it has been obtained eq.~(2.8) for the massive full gluon propagator, expressed in terms of the renormalized quantities only. It depends on the two different renormalization constants
(2.2) and (2.5), and defined up to the invariant function (2.6), which contribution at the point $q^2=M^2_g$ vanishes. The values of the massive full gluon propagator
at zero and at the point  $q^2=M^2_g$ are exactly and gauge-independently defined. Its PT asymptotic has the  $\sim 1/q^2$ renormalizable behaviour at the finite gauge, see eq.~(3.7).     
For the free massive boson propagator it goes to the expression (2.12), known as one written down in the Stueckelberg  gauge~\cite{5}. It satisfies to the ST identity expressed in the generalized gauge 
(2.10), i.e., the Stueckelberg gauge is its a particular case.

The important issue of the peculiarities of the canonical gauge  $\tilde{\lambda}^{-1} = \lambda^{-1} = \infty$ for the massive full  
gluon propagator has been investigated in sections 4 and 5. Within the mass gap approach to QCD the canonical gauge should be abandoned 
because of its two unacceptable features. Firstly, it makes the gauge particle Green functions non-renormalizable in the PT limit.
Secondly, it breaks down the equivalence between the PT $q^2 \rightarrow \infty$  and the $M^2_g=0$ and ({\it vice \ versa}) limits in the longitudinal component
of the massive full gluon propagator, and thus breaks the correct transition $\xi \rightarrow \xi_0$ in the above-mentioned limits.
This equivalence is requested by the mass gap approach to QCD in order to ensure its renormalizability.
Only the finite gauge-fixing parameters can guarantee this.The important role of the ST identity (1.3) in the formulation of the self-consistency condition
is to be underlined. It is based on the comparison between the GGD and EGD formalisms introduced in section 1.
Due to it the canonical gauge is shown to be inconsistent in the massive YM theory, and thus the GGD/EGD formalism is important in this theory.

It has been explicitly shown that no any physical meaning can be assigned to the massive full gauge particle propagator with exactly defined gluon pole mass. 
This happens because it can not determine its mass-shell, which exactly proven in section 5. That is why throughout our paper we do not call the pole position $q^2= M^2_g$
as the mass-shell condition for the massive full gluon propagator (2.8) and its free massive counterpart (2.2). However, the
existence of the massive solution shows the general possibility for a massless vector particle to acquire mass dynamically in the
vacuum. Within our approach there is no need to introduce to the theory some extra massive degrees of freedom or extract them from the theory by some other way.
At the same time, the massive gluon propagator cannot be the mass-shell object within the mass gap approach to QCD, and thus the gluon cannot appear in the physical spectrum
(confinement of massive gluons).The massive gluons may only exist in the true QCD ground state or inside hadrons, treated as its colour-singlet massive excitations.
The massive solution is confining but it does not possesses the AF regime~\cite{14,15} since in this limit it behaves simply like $\sim 1/q^2$ as mentioned above, i.e., no any  
scale breaking effect takes place (see appendix B as well).

Let us make a few preliminary remarks. It seems to us that our general conclusions are in agreement with the well-known Elitzur's theorem~\cite{17} and for its detailed discussion see~\cite{18}.
Despite the local gauge symmetry of the QCD Lagrangian is broken in its ground state within the mass gap approach to QCD, nevertheless, the massive solution of the gluon SD equation 
cannot appear in the physical spectrum, i.e., the confinement of the massive gluons remains the gauge-invariant phenomenon as it needs be. 
In more detail this theorem will be discussed in the third paper of our series.

Concluding, we have explicitly shown that the tadpole term survives the NP renormalization program, indeed. It becomes the exactly defined gluon pole mass. No any
assumptions/truncations/approximations have been made in achieving these results.
The renormalized massive full gluon propagator has been exactly derived up to the invariant function, vanishing at the pole position.  Its properties make it suitable 
for the lattice calculations in Appendix A.  In Appendix B the PT full gluon propagator has been investigated in some details, showing explicitly that gluons always remain massless in this case. 
In other words, the PT cannot generate a mass. So that the first of the two new solutions, namely massive one, of the gluon SD eq.~(1.1) has been completed in this work.
The second new solution of eq.~(1.1), namely singular one, will be completed in the third paper of the series. 

\section*{acknowledgments}

The authors are grateful to P. Forg\'{a}cs, J. Nyiri, T.S. Bir\'{o}, M. Vas\'{u}th, Gy. Kluge
for useful suggestions, remarks, discussions and help.
The work was supported by the Hungarian National Research, Development and Innovation Office (NKFIH) under the contract numbers OTKA K135515 and NKFIH 2019-2.1.11-T\'ET-2019-00078, 2019-2.1.11-T\'ET-2019-00050, the Wigner Scientific Computing Laboratory.

\appendix

\section{Euclidean metric for the lattice calculation}
%\label{euclid}

Let us begin with pointing out that the gluon pole mass is different from the effective gluon mass.
The former one is exactly defined by the relation (2.1) in a gauge-invariant way, while the latter one is to be extracted from
the effective gluon mass function~\cite{19}. The numerical value of
$M^2_g$ can be fixed by lattice QCD simulations, for example such as in~\cite{20}, but within our approach.
For this purpose, let us write down the massive full gluon propagator (2.8) in Euclidean metric by making the substitution $q^2 \rightarrow - q^2$
in (2.8), so that $q^2=q^2_0 + \bf q^2$ becomes from now on, and thus the pole position is defined as $q^2= - M^2_g$. Then one obtains
\begin{equation}
D^R_{\mu\nu}(q) = { i Z_3  {\tilde Z}^{-1}_3 \over (q^2 + M^2_g)[ 1 + Z_3 {\tilde \Pi}(q^2)]} T_{\mu\nu}(q) + i { q_{\mu}q_{\nu} \over q^2} { {\tilde\lambda}^{-1} \over (q^2 + {\tilde\lambda}^{-1} M_g^2)},
\end{equation}
where now $T_{\mu\nu}(q) = \delta_{\mu\nu} - L_{\mu\nu}(q)$. The expansion for the invariant function ${\tilde \Pi}(q^2)$ then looks like
\begin{equation}
{\tilde \Pi}(q^2) = (q^2 + M^2_g) \left[ \Pi'(-M^2_g) + q^2 \Pi''(-M^2_g) + q^2(q^2 + M^2_g) \Pi'''(-M^2_g) + ... \right],
\end{equation}
while for the corresponding NP MP renormalization constants one arrives at
\begin{equation}
{\tilde Z}_3^{-1}= [ 1  +  \Pi(-M^2_g)], \quad Z_3^{-1} = {\tilde Z}_3^{-1} - M^2_g \Pi'(-M^2_g).
\end{equation}
It is instructive to repeat the derivation of eq.~(A.1) at $q^2=0$ in this metric  as it has been done in section 3. Omitting rather simple steps, one finally obtains
equation, coinciding completely with eq.~(3.2), as it needs be.

It will be useful to introduce
the following notation as $\Pi^{(n)}(-M^2_g) = a_n, \quad n = 0,1,2,3$, where $a_n$ are the arbitrary constants of the corresponding dimensions.
Then eq.~(A.1), after doing some algebra, will be expressed in terms of these constants only, namely
\begin{equation}
iD^R_{\mu\nu}(q) = - { (1 + a_0) T_{\mu\nu}(q) \over (q^2 + M^2_g) D(q^2)}
 - { q_{\mu}q_{\nu} \over q^2} { {\tilde\lambda}^{-1} \over (q^2 + {\tilde\lambda}^{-1} M_g^2)},
\end{equation}
where

\begin{equation}
D (q^2) = [ (1 + a_0) + a_1 q^2 + (q^2 + M^2_g)q^2(a_2 + a_3 (q^2 + M^2_g) + ...)].
\end{equation}
The higher order terms ($a_4$ and higher) in its denominator can be
suppressed in the NP region $ -M^2_g \leq q^2 \leq 0$ we are mainly interested in. We are not interested in the PT tail of the massive full 
gluon propagator, i.e., in the perturbative region $ q^2 \leq -M^2_g$. Its massive free counterpart is

\begin{equation}
iD^0_{\mu\nu}(q) =  - { 1 \over (q^2 + M^2_g)} \left[ \delta_{\mu\nu} - (1 - {\tilde\lambda}^{-1})
{q_{\mu}q_{\nu} \over (q^2 + {\tilde\lambda}^{-1} M^2_g)} \right],
\end{equation}
Compare this equation with eq.~(2.12) in Minkowski metric. However, let us remind that the values at zero $q^2=0$ and at
$q^2= - M^2_g$ are exactly and gauge-invariantly defined. Also our estimation~\cite{11} of the scale of the NP dynamics in QCD was about $(0.5-0.6)$ GeV.
The pole mass of a single gluon is expected to be of the same order of magnitude. Note that the value of $m_g \approx 550$ MeV has been calculated in~\cite{20} 
(and see references therein for other determinations).

\hspace{2mm}

\subsection*{Remarks on the gluon pole mass}

\hspace{2mm}

The massive gluons with non-zero pole masses, which are exactly defined in (2.1), may exist in the
QCD ground state and inside hadrons along with their other massive and massless counterparts. They present new degrees of freedom of the massive gluon field configurations in the QCD vacuum.
We call such a solution for the massive full gluon propagator as NP massive (since the gluon pole mass is of purely NP dynamical origin, as described above).
Due to its asymptotic properties, the gluon pole mass contribution is to be neglected in the PT $q^2 \rightarrow \infty$ limit.
This means that in the experiments at high energies involving the strongly-interacted particles, it is impossible to fix the gluon pole mass $M_g^2$ as stated in section 3.
Apparently, "experimentally" the gluon pole mass can be fixed by the lattice simulations, as
formulated just above. The dynamically generated gluon pole mass looks like very similar to the current mass of a free quark.
The principle difference is that now we know how it may appear in the full gluon propagator.
The dynamical source of the current quark mass is still remains unknown, though its term is compatible with the $SU(3)$ colour gauge invariance of the QCD 
Lagrangian, while the gluon pole mass term does not. In the mass gap approach to QCD and its ground state the gauge symmetry of the QCD Lagrangian is not a symmetry of its ground state. 
However, let us remind that it is not a problem, see our fist paper~\cite{1} as well as the third paper of our series and our present investigation showing that the absence of the mass-shell 
in the massive full gluon propagator is a gauge-invariant effect.

There are no any doubts that inside hadrons and nuclei quarks can interact with each other by exchange of not only massless gluons but their massive counterparts as well. 
The interaction between heavy quarks due to the exchange of massive gluons is described by the Yukawa-type potential $V(r) \sim (1/r) \exp(-M_gr)$, though it is not confining quarks, it is strong but short-range (at $M^2_g=0$ it becomes  the Coulomb-type one, namely $V(r) \sim (1/r)$,  and thus describes the interaction between quarks due to exchange of the massless gluons).
This means that one can 'see' the massive gluons inside
the strongly-interacting particles (hadrons) only. But one cannot 'see' them inside hadrons at short distances ($q^2 \rightarrow \infty$), as pointed out just above.
The massive gluons with exactly defined the gluon pole masses will additionally contribute to the quarks effective masses (properly defined) via the quark SD equation, making them much more different
from the current quark masses. This knowledge may substantially improve our understanding of the dynamical structure of
the QCD ground state, hadrons and nuclei internal structures and their properties, such as masses~\cite{21}, nucleon
and glue spins, etc.~\cite{22,23}. The existence of massive gluons can provide more information about the ordinary nuclear matter in the interior
of compact (neutron) stars and their merger~\cite{24}. The gluons with exactly defined pole masses may also play important role in the creating of the different phases of QCD Matter at high temperature and density~\cite{25}. They will provide the gluon degrees of freedom, but
different from the quasi-particles which also show up as effective poles (depending on the temperature) in the gluon propagators, calculated by the thermal lattice QCD~\cite{26}. 
The massive gluons should be included into the YM NP equation of states, e.g. such as derived in~\cite{27}. One can easily imagine the glueballs as bound states of the two or three gluons with pole masses and not only consisting of the gluons with effective gluon masses or massless gluons~\cite{28}.

Let us make a few more brief remarks about gluon masses.
The existence of the mass gap has been studied within lattice QCD in recent paper~\cite{29}. However, its mass gap cannot be identified with the gluon pole mass, rather then with the scale breaking?
The pioneering work~\cite{30}, where the gluon mass has been introduced into the continuum QCD, is to be acknowledged.
The discussion of other approaches and models in the different simplifications and gauges for
a possible existence of the gluon mass, for example such as Curci\,--\,Ferrari~\cite{31}, Gribov\,--\,Zwanziger~\cite{32,33}, the effective gluon
mass function~\cite{19,34}, lattice simulations~\cite{20} (and see references therein), dual QCD~\cite{35}, LCO formalism to calculate
the gluon mass~\cite{36} and many others up to present days can be found in the recent reviews~\cite{34,37,38}.

\hspace{2mm}

\section{The PT massless full gluon propagator}

\hspace{2mm}

The PT full gluon propagator,  derived in~\cite{1},  in Minkowski metric is

\begin{equation}
D^{PT}_{\mu\nu}(q) =  {- i T_{\mu\nu}(q) \over q^2[ 1 + \Pi(q^2) ]}  - i  \lambda^{-1} L_{\mu\nu}(q) { 1 \over q^2 }.
\end{equation}
It can be obtained from eq.~(1.1) by putting there formally $ M^2 = \Delta^2_t(D) = 0$. 
This means that the invariant function now is defined as $\Pi(q^2) = \Pi^s(q^2; D^{PT})$, and thus it depends on $D^{PT}$ and not on $D$ like in eq.~(1.1).
However, as it follows from~\cite{1}, nothing is changed in its asymptotic properties, i.e., it remains  a regular function of its argument at zero and only logarithmic divergent
in the PT $q^2 \rightarrow \infty$ limit.

It is instructive first to explicitly show that this expression for the PT full gluon propagator follows from eq.~(2.8), and all the relations connected to it,  at $M_g^2=0$.
The pole position now goes to $q^2=0$, i.e., eq.~(B.1) describes the propagation of the massless regularized gluon fields in the QCD ground state.
This means that we are putting the treatment of the tadpole term on the same footing as the quark and gluon regularized
constants, i.e., all the quadratically divergent but regularized scale parameters are to be disregarded and thus removed from the theory on the general 
mathematical basis (they omission is not a prescription as it has been proven in~\cite{1}).
The exact gauge symmetry of the QCD Lagrangian is restored in its ground state in this case, and because of this effect eq.~(B.1) has been conventionally 
called the PT one.  It is valid in the whole gluon momentum range, but the coupling constant remains strong apart from the AF regime~\cite{14,15}. 
 
So letting $M_g^2=0$ at finite ${\tilde\lambda}^{-1}$ in eqs.~(2.8)-(2.10), one obtains
\begin{equation}
D^R_{\mu\nu}(q) = { - i  T_{\mu\nu}(q) \over q^2 \left[ 1 + Z_3 (0) \Pi^r(q^2)\right] } 
- i {\tilde\lambda}^{-1} L_{\mu\nu} (q) { 1 \over q^2},
\end{equation}
where $\tilde\lambda = \lambda Z_3(0)$ and the invariant function $\Pi^r(q^2) = {\tilde \Pi}(q^2; M_g^2=0)$ and thus is defined by the expansion (3.6) in this limit, so it is
\begin{equation}
\Pi^r(q^2) = \Pi(q^2) - \Pi(0) = q^2 \Pi'(0) + q^4 \Pi''(0) + q^6 \Pi'''(0) + ... .
\end{equation}
The NP MP renormalization constants $Z_3(0)$ and ${\tilde Z}_3(0)$, given by the relations (2.2) and (2.5), respectively, coincides with each other at $M_g^2=0$

\begin{equation}
Z_3(0)=Z_3(M^2_g=0) = {\tilde Z}_3(M^2_g=0) = {1 \over 1 + \Pi(0)},
\end{equation}
as it should be in the massless case, indeed.

Using these relations and definitions (2.9), from eq.~(B.2) one obtains

\begin{equation}
D_{\mu\nu}(q) = { - i T_{\mu\nu}(q) \over q^2 \left[ 1 + \Pi(q^2)\right] } 
- i {\lambda}^{-1} L_{\mu\nu} (q) { 1 \over q^2},
\end{equation}
where the initial invariant function now is given by the Taylor expansion as follows:
$\Pi(q^2) = \sum_{n=0}^{\infty} (q^2)^n \Pi^{(n)}(0)$,
in agreement with the expansion (2.3), as expected, and where $\Pi^{(n)}(0)$ denotes its corresponding derivatives at the point $q^2=0$.
Eq.~(B.5) is nothing else but eq.~(B.1), denoted as  $D^{PT}_{\mu\nu}(q)$. 

Putting formally $\Pi(q^2) = 0$ in eq.~(B.5), for the free massless vector boson propagator one obtains eq.~(3.7),
so that from eqs.~(B.5) and (3.7) one arrives at
$q_{\mu}q_{\nu} D_{\mu\nu}(q) =  q_{\mu}q_{\nu} D^0_{\mu\nu}(q) = - i {\lambda}^{-1}$.
Eq.~(B.2) is the renormalized version of eq.~(B.5) ,  and hence of the PT massless full gluon propagator (B.1).

%\section*{References}

{}


\begin{thebibliography}{}
\bibitem{1}
   V. Gogokhia, G.G. Barnafoldi, The Mass Gap Approach to QCD. I. The true gauge and dynamical structures of its ground state, arXiv:2301.04561v5  
\bibitem{2}
   W. Marciano, H. Pagels, Quantum Chromodynamics, \\ Phys. Rep. C, 36 (1978) 137.
\bibitem{3}
   J. C. Taylor, Ward identities and charge of the Yang-Mills field, Nucl. Phys. B, 33 (1971) 436.
\bibitem{4}
   A. A. Slavnov, Ward identities in Gauge theories, Theor. Math. Phys., 10 (1972) 153.
\bibitem{5}
   C. Itzykson, J.-B. Zuber, Quantum Field Theory (McGraw-Hill Book Company, 1984).
\bibitem{6}
   G. 't Hooft, Renormalizable Lagrangians for Massive Yang-Mills Fields,  Nucl. Phys. B, 35 (1971) 167.
\bibitem{7}
   T.D. Lee, C.N. Yang,  Theory of Charged Vector Mesons Interacting with the Electromagnetic Field, Phys. Rev., 128 (1962) 885.
\bibitem{8}
   K. Fujikawa, B.W. Lee, A.I. Sanda,  Generalized Renormalizable gauge Formulation of Spontaneously Broken Gauge Theories, Phys. Rev. D, 6 (1972) 2923.
\bibitem{9}
   B.W. Lee, J. Zinn-Justin,  Spontaneously Broken Gauge Symmetries. IV. General Gauge Formulation, Phys. Rev. D, 7 (1973) 1049.
\bibitem{10}
   A. Jaffe, E. Witten, Yang\,--\,Mills Existence and Mass Gap, \\
   $http://www.claymath.org/prize-problems/, \
   http://www.arthurjaffe.com$ \ .
\bibitem{11}
    V. Gogokhia, G.G. Barnaf\"oldi, The Mass Gap and its Applications (World Scientific, 2013).
\bibitem{12}
    G.A. Korn, T.M. Korn, Mathematical Handbook 
    
    (McGraw-Hill Book Company, 1968).  
\bibitem{13}
     J.B. Conway, Functions of One Complex Variable (Springer, 1978).
\bibitem{14} 
     D. J. Gross, F. Wilczek, Ultraviolet behaviour of non-abelian gauge theories, Phys. Rev. Lett., 30 (1973) 1343.
\bibitem{15} 
    H.D. Politzer, Reliable perturbative results for strong interactions?, Phys. Rev. Lett., 30 (1973) 134
\bibitem{16}
     Y. L. Dokshitzer, D.E. Kharzeev, The Gribov conception of Quantum Chromodynamics, Ann. Nucl. Part. Sci., 54 (2004) 487
\bibitem{17}
     S. Elitzur, Impossibility of spontaneously breaking local symmetries, 
     
     Phys. Rev. D, 12 (1975) 3978.
\bibitem{18}
   J. Greensite, An Introduction to the Confinement Problem
   
   (Springer, 2 ed., 2020).
\bibitem{19}
   B. Holdom, Soft asymptotics with mass gap, 
   
   Phys. Lett. B, 728 (2014) 467.
\bibitem{20}
   A. Cucchieri, D. Dudal, T. Mendes, N. Vandersickel, Modelling the Gluon Propagator in Landau Gauge: Lattice Estimates of pole masses and Damensional-Two Condensates, Phys. Rev. D, 85 (2012) 094513.
\bibitem{21}
    M. Tanabashi, et. al., Particle Data Group, Phys. Rev. D, 98 (2018).
\bibitem{22}
    C. A. Aidala, S.D. Bass, D. Hasch, G.K. Mallot,The spin structure of the nucleon, Rev. Mod. Phys., 85 (2013) 655.
\bibitem{23} 
    Yi-Bo Yang et. al., Glue spin and helicity in the proton from lattice QCD, Phys. Rev. Lett., 118 (2017) 102001-1.
\bibitem{24}
    A. Jakov\'ac, G.G. Barnaf\''oldi, P. P\'osfay, Effect of quantum fluctuations in the high-energy cold nuclear equation of state and in compact star observations, Phys. Rev. C, 97 (2018) 025803.
\bibitem{25}    
    {\it Quark Matter 2022}, Proc. of the XXIV Inter. Conf. on Ultra-Relativistic Nucleus Collisions, ed. by P. Braun-Munziger, B. Friman, J. Stachel, Nucl. Phys. A, 931 (2014) 1. 
\bibitem{26}
    P. Petreczky, F. Karsch, E. Laermann, S. Stickan, I. Wetzorke, Temporal quark and gluon propagators: Measuring the quasiparticle masses, Nucl. Phys. Proc. Suppl., 106 (2002) 513.    
\bibitem{27}
    V. Gogokhia, M. Vasuth, The non-perturbative analytical equation of state for the gluon matter, J. Phys. G: Nucl. Part. Phys., 37 (2010) 075015.
\bibitem{28}
    C. Bernard, Nonte Carlo Evaluation of the Effective Gluon Mass, 
    
    Phys. Lett. B, 108 (1982) 431.                     
\bibitem{29}
    R. Vilela-Mendes, A consistent measure for lattice Yang-Mills, 
    
    Int. J. Mod. Phys. A, 32 (2017) 1750016.
\bibitem{30}    
    J. M. Cornwall, Dynamical mass generation in continuum quantum 
    
    chromodynamics, Phys. Rev. D, 26 (1982) 1453.
\bibitem{31}
    G. Cursi, R. Ferrari, The unitary problem and the zero-mass limit for a model of massive Yang-Mills theory, Nuovo Cim., A35 (1976) 1.
\bibitem{32}
     V.N. Gribov, Quantization of non-Abelian gauge theories, 
     
     Nucl. Phys. B, 139 (1978) 1.
\bibitem{33}
    D. Zwanziger, Action from Gribov horizon, 
    
    Nucl. Phys. B, 321 (1989) 591.
\bibitem{34}
    M.Q. Huber, Nonperturbative properties of Yang-Mills theories, 
    
    Phys. Rep., 879 (2020) 1-92.
\bibitem{35}
    M. Baker, J.S. Ball, F. Zachariasen, Dual QCD: a review, 
    
    Phys. Rep., 209 (1991) 73.
\bibitem{36}
    J.A. Gracey, Two loop gluon mass from the LCO formalism, 
    
    Eur. Phys. J. C, 39 (2005) 61.
\bibitem{37} 
    M. N. Ferreira, J. Papavassiliou, Gauge Sector Dynamics in QCD, 
    
    Particles 2023, 1, 1-57.
\bibitem{38}
    J. Papavassiliou, Emergence of mass in the gauge sector of QCD, 
    
    Chinese Phys. C, 46 (2022) 112001.                
\end{thebibliography}
\end{document}